\documentclass[a4paper]{jpconf}
\usepackage[utf8x]{inputenc}
\usepackage[english]{babel}
\usepackage{amsmath}
\usepackage{multicol}
\usepackage{hyperref}
\usepackage{color}
\usepackage{graphicx}

\begin{document}
%
%
\title{Numerical determination of the CFT central charge in the site-diluted Ising model}
  \author{P A Belov$^{1}$, A A Nazarov$^{1}$ and A O Sorokin$^{2}$}
  \address{$^{1}$Department of Physics, St. Petersburg State University, Ulyanovskaya 1, 198504 St.~Petersburg, Russia\\ $^{2}$Petersburg Nuclear Physics Institute NRC Kurchatov Institute, Orlova roscha 1, Gatchina, 188300 Leningrad region, Russia}
\ead{antonnaz@gmail.com}
%
%
%
\begin{abstract}
  We propose a new numerical method to determine the central charge of the conformal field theory models corresponding to the 2D lattice models.
  In this method, the free energy of the lattice model on the torus is calculated by the Wang-Landau algorithm and then the central charge is obtained from a free energy scaling with respect to the torus radii.
The method is applied for determination of the central charge in the site-diluted Ising model.
\end{abstract}

\section{Introduction}

Spin models play an important role in statistical physics.
They describe real systems, namely magnets, their non-trivial properties and
phenomena, especially at a
critical point.
These properties can be studied by various analytical and numerical approaches, so the spin models are the proper frameworks for their development and verification.

It is well known~\cite{Polyakov:1970xd} that at a critical point a spin system has the conformal symmetry,
so the conformal field theory (CFT) is of special importance in this case.
In two dimensions (2D), CFT is
especially powerful since it possess a rich symmetry represented by the infinite-dimensional
Virasoro algebra~\cite{BPZ}:
\begin{equation*}
  \label{eq:2}
\left[L_n,L_m\right]=(n-m)L_{n+m}+\frac{c}{12}(n^3-n)\delta_{n+m,0},
\end{equation*}
where $L_{n,m}$ are the generators of the Lie algebra with a central extension,
$\delta_{n+m,0}$ is the Kronecker delta. The appearance of the central charge, $c$, is a result of the conformal anomaly. It arises in quantum theory in the two-point correlation
function of the stress-energy tensor~\cite{CFTbook}.
Models of CFT have a small number of parameters, namely the central charge and
conformal weights of the primary fields, $h_{i}$, which are the eigenvalues of the operator $L_{0}$.
These models allow one to obtain
analytically or numerically a lot of observables, such as multi-point
correlation functions, to study different boundary conditions, behavior in certain geometries and
various perturbations of CFT models~\cite{CFTbook,cardy1986effect,cardy1984conformal,fateev1990conformal}.

For most famous lattice models the CFT counterparts are analytically known.
For example, the Ising model corresponds to the CFT characterised by
$c=1/2$~\cite{BPZ}.
There are some
models like the $q$-state Potts model~\cite{PottsReview} and the tricritical Ising
model~\cite{blume1966theory,capel1966possibility} whose lattice counterparts and field parameters
are also known.

However, there are models whose counterparts as well as the CFT parameters are unknown. A remarkable
example is the site-diluted Ising model~\cite{ballesteros1997ising} which implies presence of
magnetic clusters in the whole system. This model is usually characterized by two parameters:
temperature $T$ (or exchange energy $J$) and the probability $p$ of a lattice site to be magnetic. The
attempts to describe the critical behavior of this system have a long history and the agreement has
not been achieved so far. Here, we can mention papers of Dotsenko and
Shalaev~\cite{dotsenko1983critical,shalaev1984correlation} who theoretically obtained the critical
behavior of correlation length, magnetic suspectibility and heat capacity.
There are also many numerical studies of the critical behavior~\cite{andreichenko1990monte,kim1994critical,najafi2016monte}, some of them contradict with the theorerical results.
For example, the recent paper by Najafi~\cite{najafi2016monte} studies the 2D lattice site-diluted Ising model
at the critical point.
In this study, the simulated domain interfaces were explained by the Schramm-Loewner evolution~\cite{schramm2000scaling}.
As a result,
a non-linear behavior of the central charge $1/2 \le c \le 1$ for $p=0.8-1.0$ was obtained from its
relation to the diffusion parameter \cite{bauer2002sle,bauer2003conformal}.
So, Ref.~\cite{najafi2016monte} claims that the site-diluted
Ising model is different from the pure Ising model.


In this report, we present a new method to obtain the central charge from Monte Carlo (MC) simulations. Although the
classical MC techniques~\cite{landau2014guide} do not allow one to extract the free energy, we apply the recently developed Wang-Landau algorithm~\cite{wang2001efficient} to calculate
the free energy of a lattice model on a torus. The central charge is obtained from the free energy scaling with respect to the torus radii. 
Our method is universal and can be applied to different lattice models.
We illustrate the applicability of the method by the site-diluted Ising model.
For this model we calculate the central charge, $c$, as a function of
probability, $p$, for $p=0.8-1.0$.
Our simulations show that for this model $c \approx 1/2$. This indicates that the
Ising-like behavior is preserved in the site-diluted model and the central charge is universal in
contrast to results of Ref.~\cite{najafi2016monte}.

\section{Determination of the central charge}
We study the critical behavior of the site-diluted Ising model.
Similar to the pure 2D Ising model, it is formulated on a lattice with magnetic sites at the lattice vertices.
We performed simulations on a triangular lattice, though
the critical behavior is independent of the lattice type.
Each site has a spin $s=\pm 1$ or can
be non-magnetic ($s=0$).
Sites are magnetic with the probability $p$, so the case of $p=1$
corresponds to the Ising model.
The Hamiltonian of the site-diluted model is the same as in the Ising model: 
\begin{equation}
  H=-J\sum_{\left \langle i,j \right \rangle} s_{i}s_{j},\; s_{i}=\pm 1, \mbox{or}\;0
\end{equation}
where
$\left \langle i,j \right \rangle$ denotes the sum over neighbouring sites of the lattice. 

\begin{figure}[h!]
     \centering
     \includegraphics[width=0.46\linewidth]{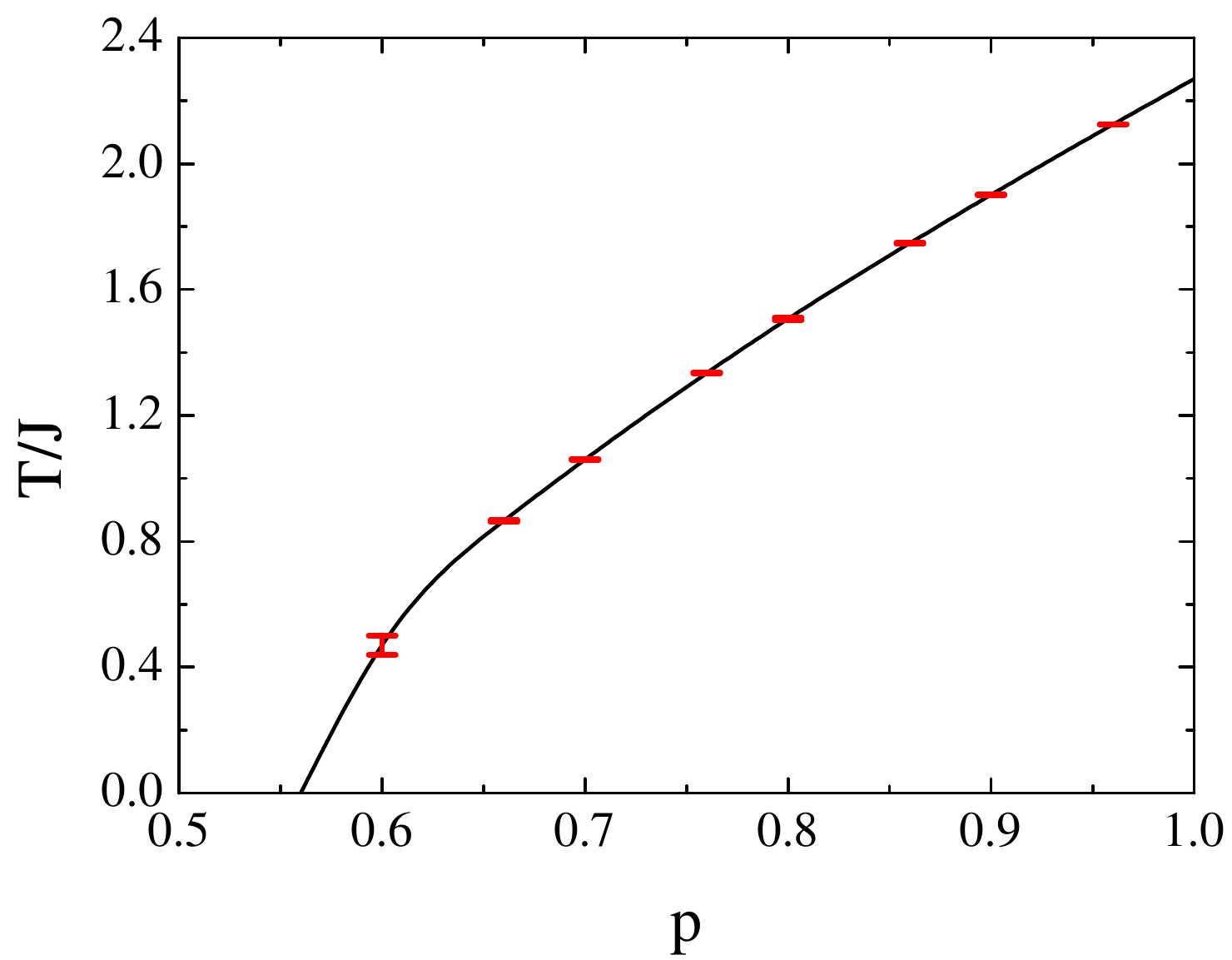}
     \ 
     \includegraphics[width=0.45\linewidth]{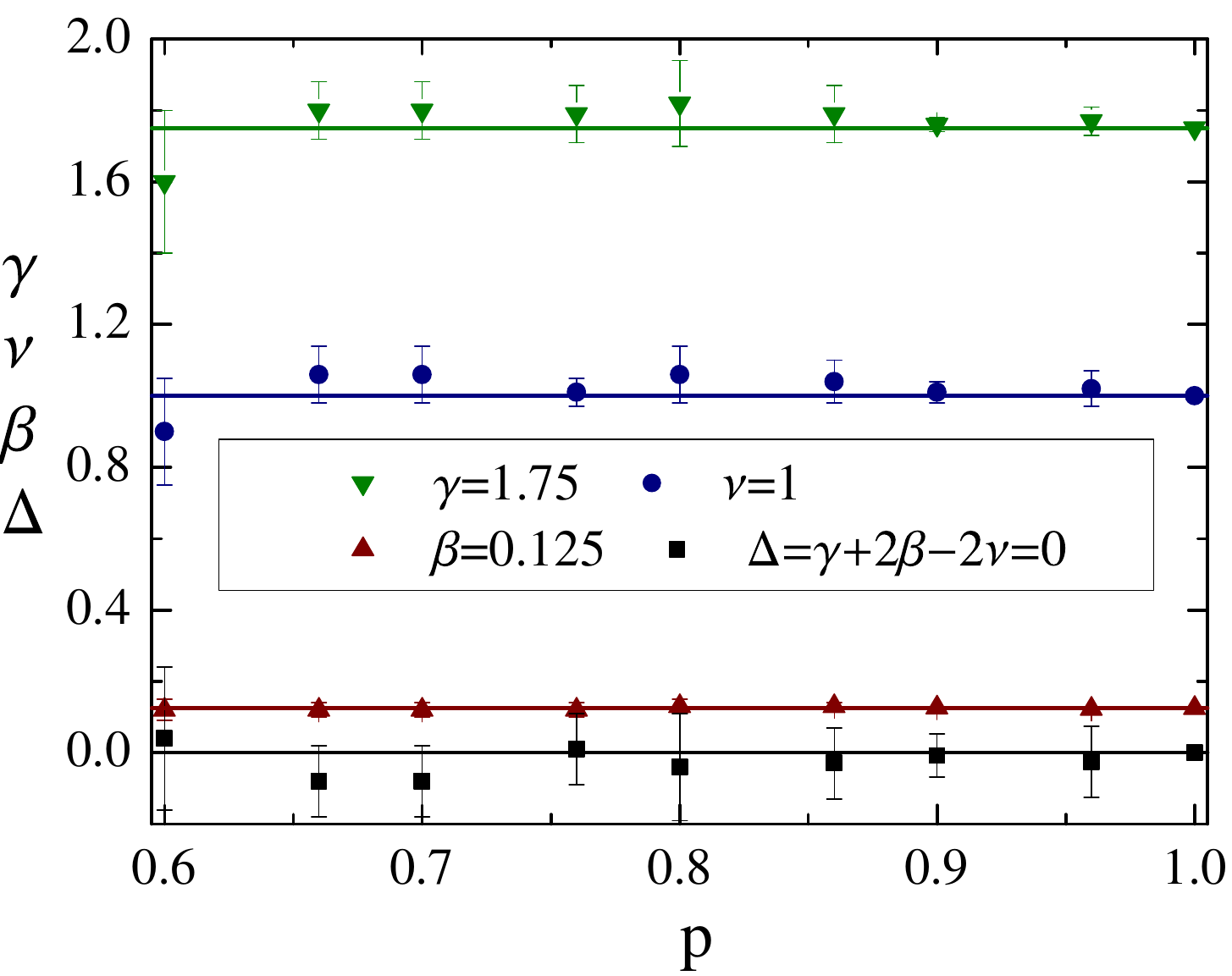}
     \caption{\label{phase} ({\it Left}) The critical line of the site-diluted Ising model obtained by the Wolff cluster algorithm; the ordered phase lies below the critical line. ({\it Right}) A comparison of the critical indices for the site-diluted and pure Ising models.}
   \end{figure}

The site-diluted Ising model has a critical point for any given value of $p$.
The critical temperature changes continuously with respect to $p$.
For example, in Figure \ref{phase} we show the phase diagram and critical indices obtained in our additional
MC simulations on a square lattice by the Wolff cluster algorithm~\cite{wolff1989collective,landau2014guide}.

%
%
%

In our method, we use the Wang-Landau algorithm~\cite{wang2001efficient} to simulate the energy distribution $g(E)$ in the partition function
\begin{equation*}
  \label{eq:4}
    Z=\sum_{[s]} e^{-\frac{E[s]}{kT}} = \sum_{E} g(E) e^{-\frac{E}{kT}}.
\end{equation*}
On each step of the algorithm, a new configuration is chosen randomly with the weight given by the accumulated value of $g(E)$.
If the new configuration is accepted, the value of $g(E)$ is increased.

We perform the simulations on $M\times N$ lattice with the periodic boundary conditions.
In this way, we obtain the partition function and free energy
\begin{equation}
  \label{eq:001}
f(M,N)=-\frac{T_{c}}{MN}\log Z(M,N)
\end{equation}
on the torus of circumferences $M$ and $N$.

The Hilbert space of CFT is a direct sum of the Virasoro algebra representations $V(c,h_{i})$: $
{\cal H}=\bigoplus_{i} V(c, h_{i})\otimes V(c,\bar h_{i} )$.
So, the partition function on the torus is given by the combination of the Virasoro characters~\cite{CFTbook}
$$
\frac{Z(q)}{Z_{0}}=\sum_{i,\bar i} {\cal M}_{i,\bar i}\chi_{i}(q)
\bar\chi_{\bar i}(\bar q),
$$
where $q=\exp{(2\pi i \tau)}$ and $\tau=iM/N$ is the modular
parameter of the torus.
If $M \gg N$, we can expand the free energy~(\ref{eq:001}) holding only leading contributions with $q^{h_{i}}$
\begin{equation}
  \label{eq:14}
  f(N,M)=f_{0} - \frac{T_{c}\pi c}{6
    N^{2}}-\frac{T_{c}}{MN}\left[\sum_{i\neq 0,\bar i\neq 0} \mathcal{M}_{i,\bar i} q^{h_{i}+\bar h_{\bar i}} \right],
\end{equation}
since the parameter $q=\exp{(-2\pi M/N)}$ is very small.
We performed simulations for different values of $M$, $N$ and obtained the dependencies of the free energy expansion~(\ref{eq:14}) on these quantities.
The extrapolation to $\frac{M}{N}\to\infty$ allowed us to extract the central charge $c$ from the second term of Eq.~(\ref{eq:14})
because the third term becomes negligible.

\section{Simulation results}       
\subsection{Ising model ($p=1$)}
\begin{figure}[t]
  \centering
  \includegraphics[width=0.41\linewidth]{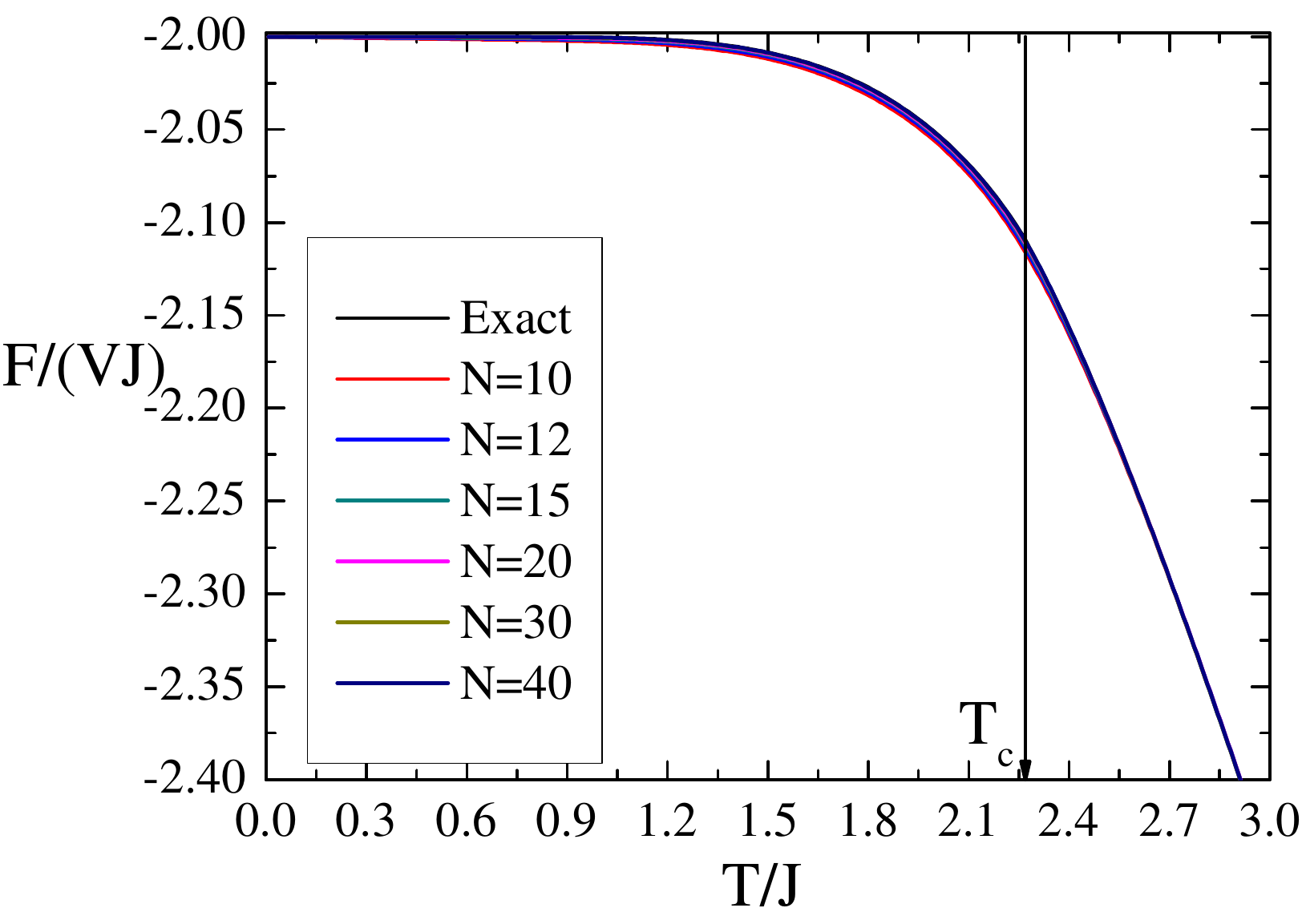}\qquad
  \includegraphics[width=0.4\linewidth]{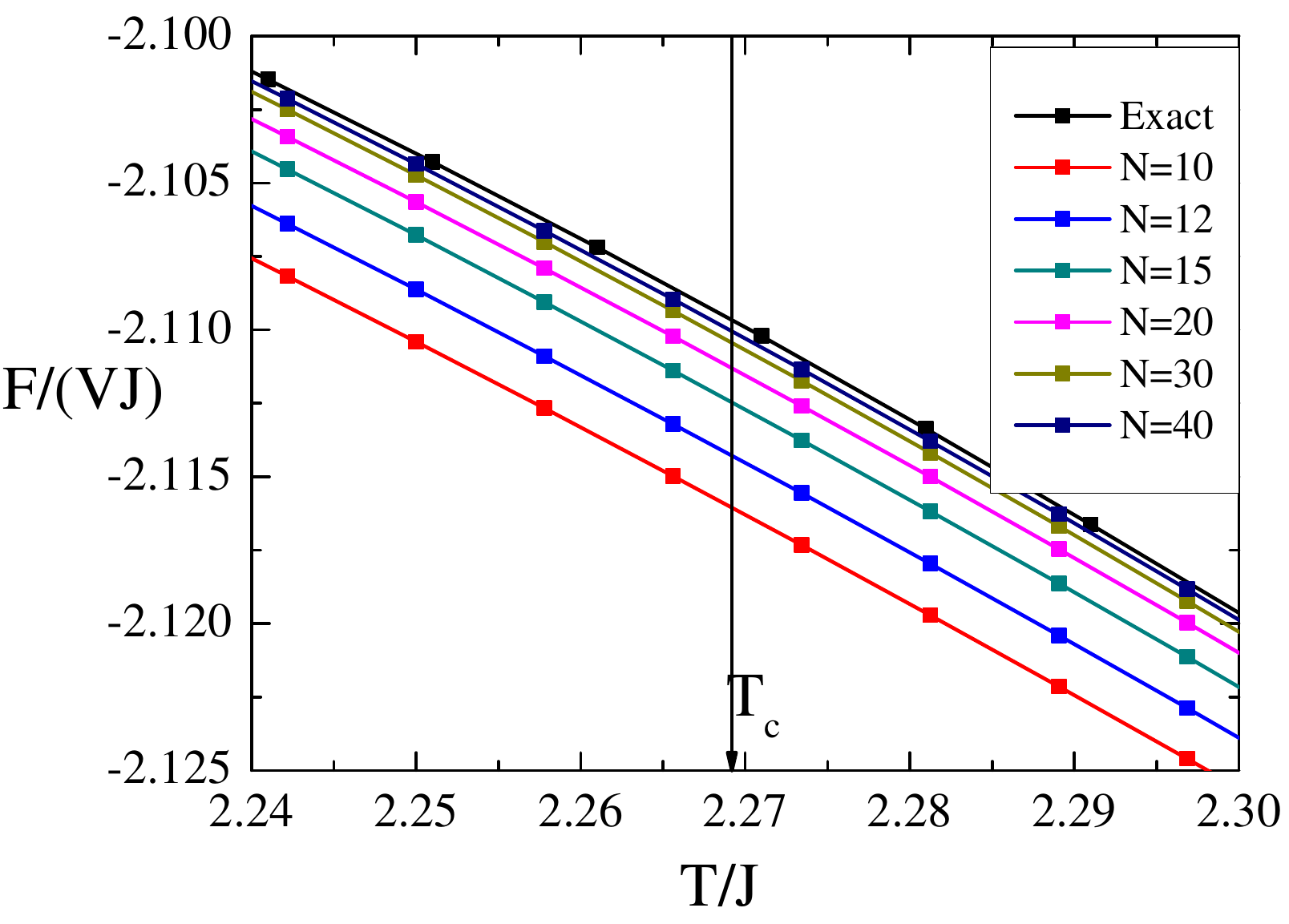}
  \caption{\label{FTIsing} A thermal dependence of the free energy for different values of $N$ in a wide range of temperatures ({\it left}) and near the critical point ({\it right}).}
\end{figure}

For the Ising model, $p=1$, we compared the free energy simulations with the exact Onsager solution~\cite{onsager1944crystal}.
It can be seen in Figure~\ref{FTIsing}, where the dependence of the free energy on the temperature is plotted.
We show this dependence for different lattice sizes $N$ together with the exact solution.

\begin{figure}[b]
  \centering
  \includegraphics[width=0.4\linewidth]{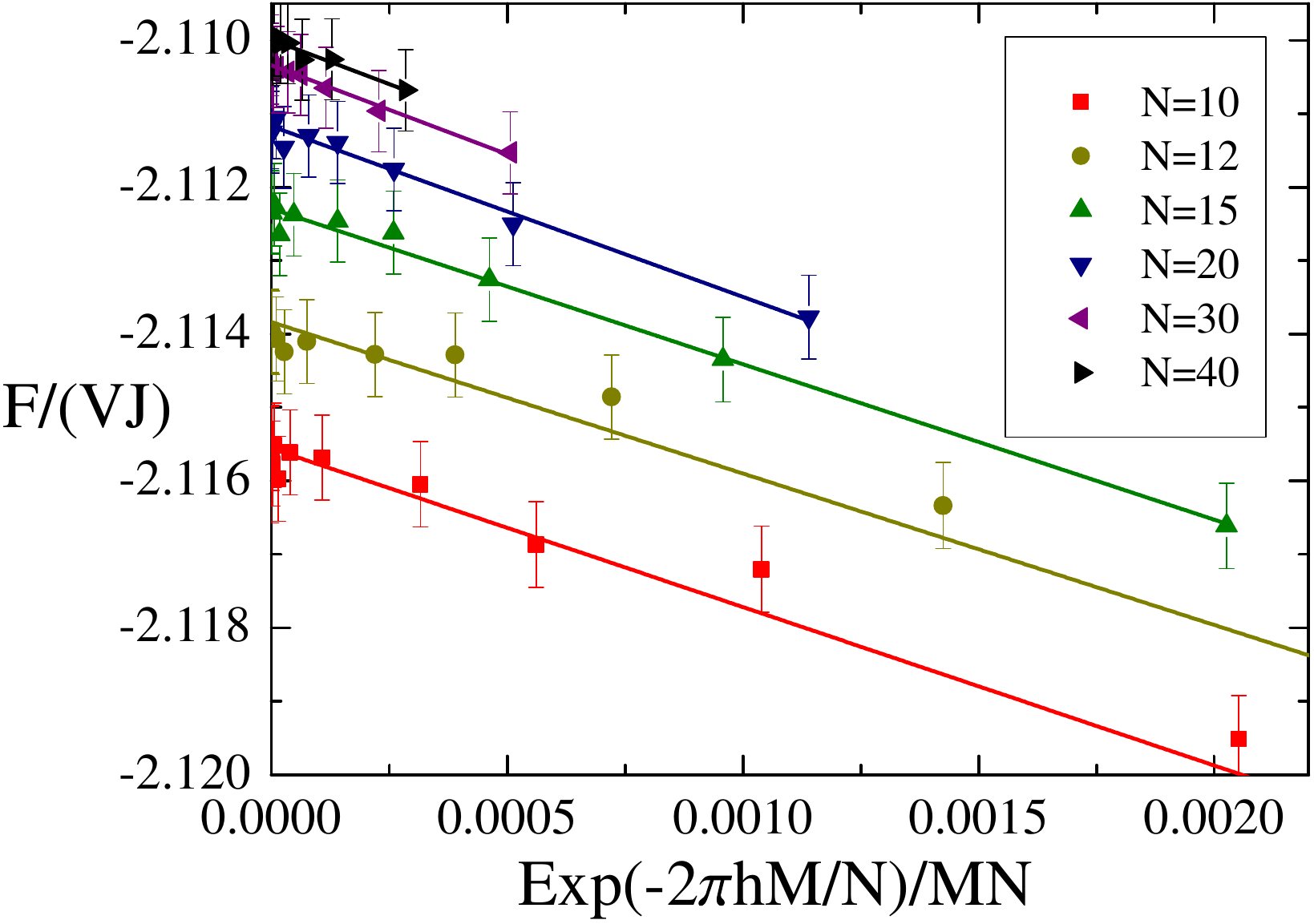}\qquad
  \includegraphics[width=0.4\linewidth]{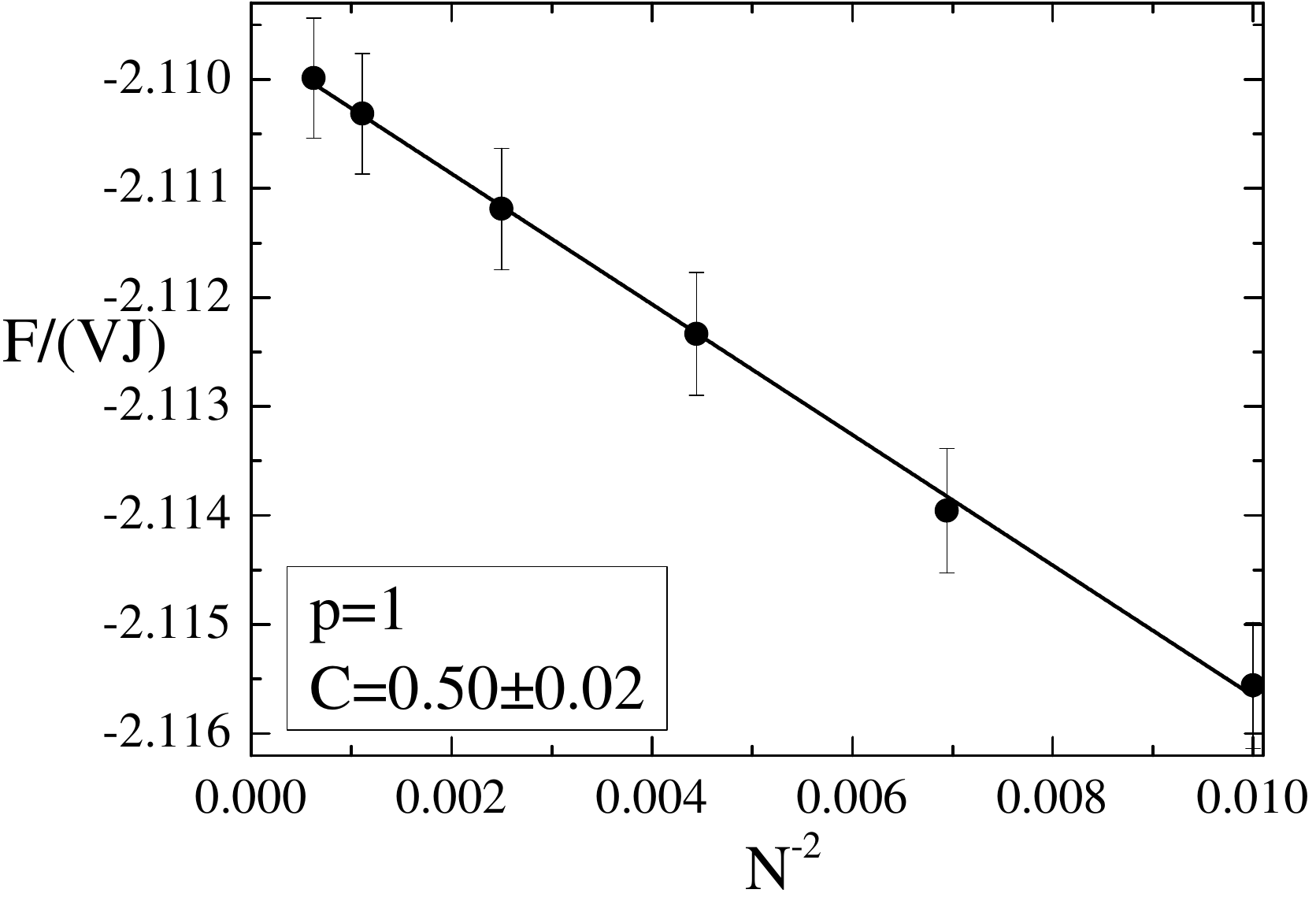}
  \caption{\label{fxnew} A finite-size scaling of the free energy $f(M,N)$ as a function of $q^{h}=e^{-2\pi h M/N}$ ({\it left}) and $f(\infty,N)$ as a function of $N^{-2}$ ({\it right}).}
\end{figure}

To determine the central charge, we first need to calculate the limit as $M \to \infty$ of the free energy $f(M,N)$ in Eq.~(\ref{eq:14}).
We obtained it by extrapolating the simulation results for different $M$ and $N$ to the limit $M\to \infty$.
For example, the dependence of the free energy on $q^{h}=\exp({2\pi h M/N})$ for the different values of $N$ is shown in Figure~\ref{fxnew}. We extrapolated the results to $q\to 0$ as $M\to\infty$ and determined $f(\infty,N)$.
Then we fitted the dependence \eqref{eq:14} and obtained the value of central charge $c=0.50\pm 0.02$,
which is in a very good agreement with the theoretical result of $c=\frac{1}{2}$.

\subsection{Site-diluted Ising model ($p<1$)}
\label{sec:site-diluted-ising}
\begin{figure}[t]
   \centering
   \includegraphics[width=0.43\linewidth]{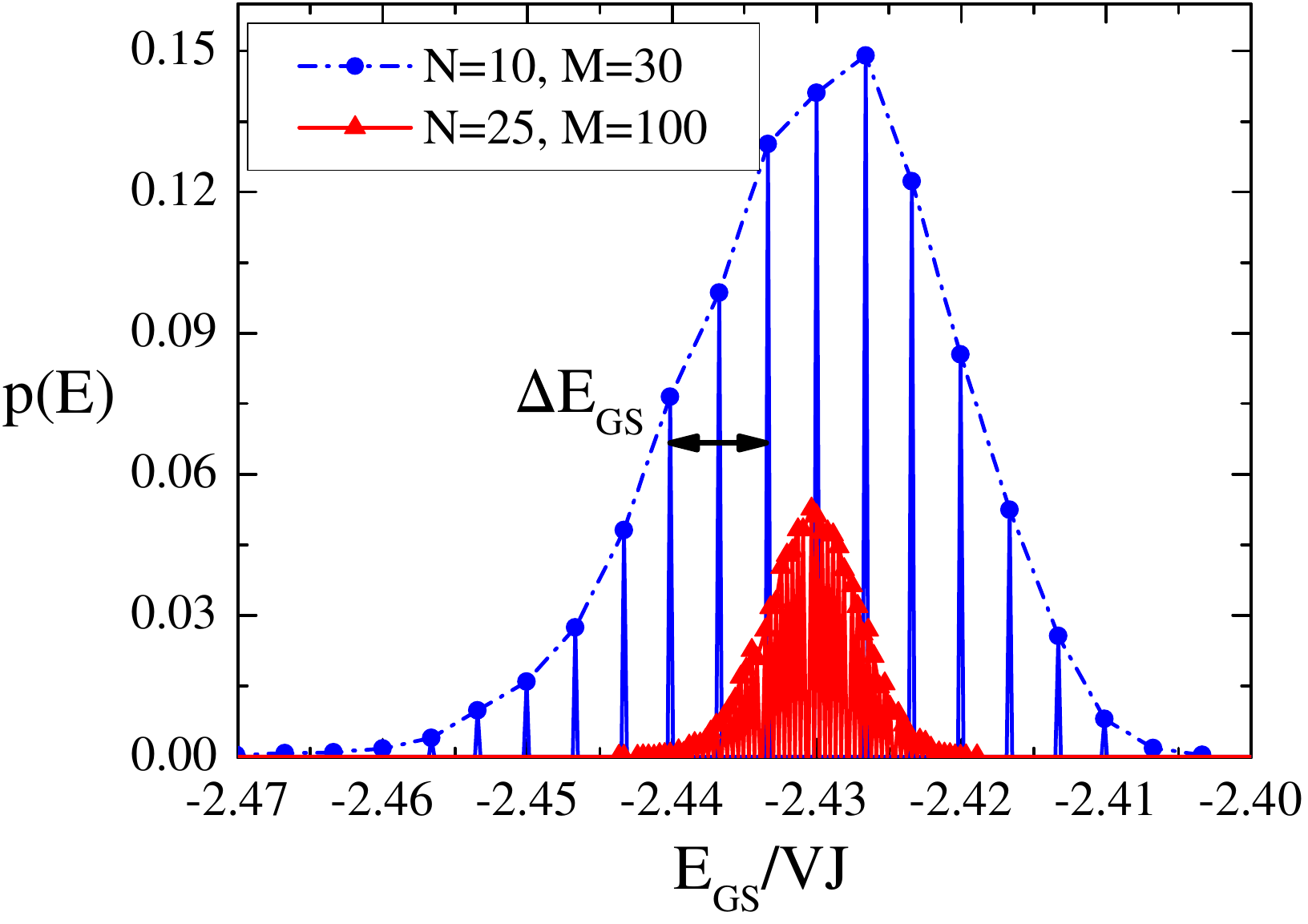}\qquad
   \includegraphics[width=0.4\linewidth]{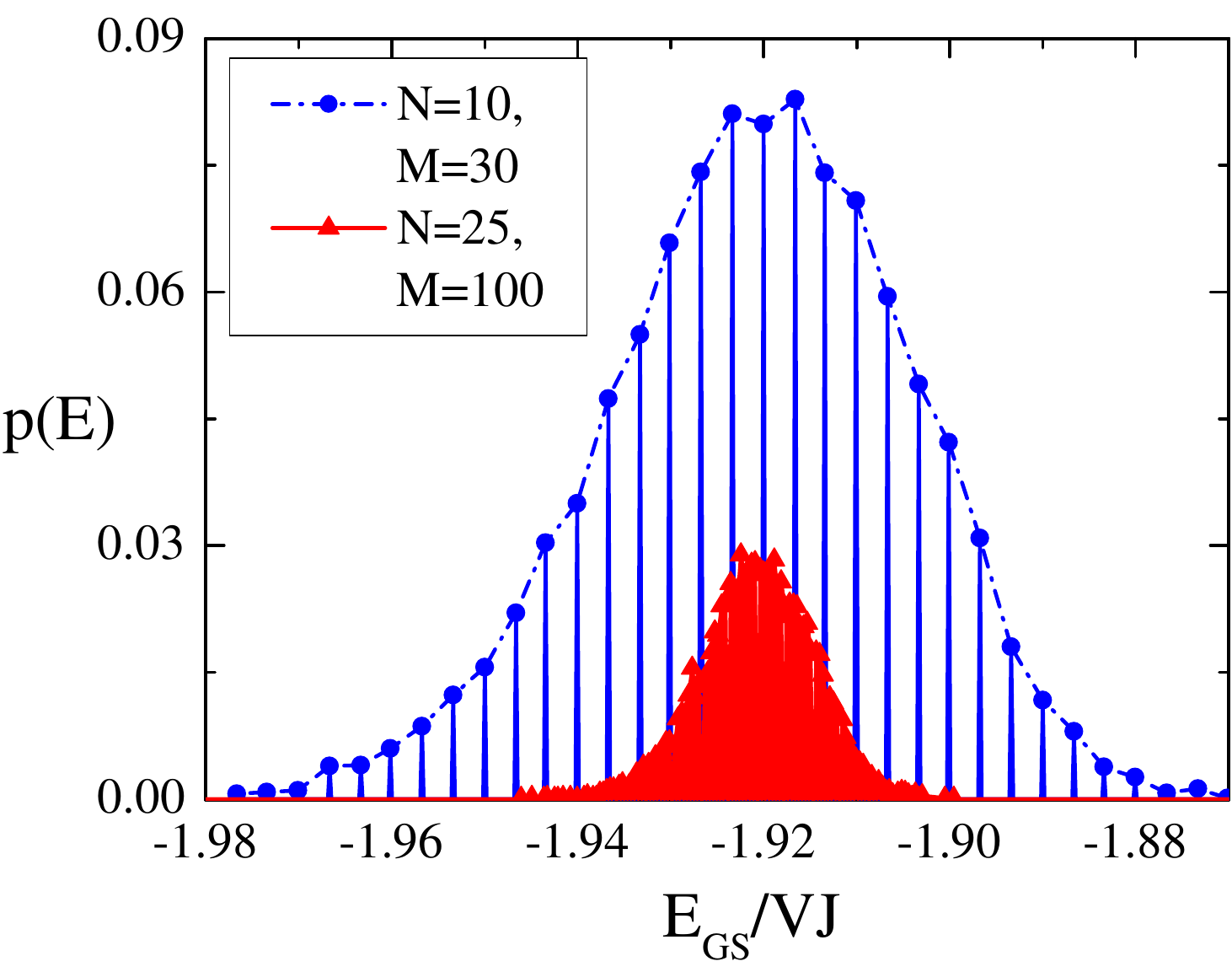}
   \caption{\label{inacc}Additional inaccuracy in an energy estimation for the site-diluted Ising model with $p=0.9$ ({\it left}) and $p=0.8$ ({\it right}).}
\end{figure}

For the site-diluted Ising model, $p<1$, we studied probabilities $p=0.8,\;0.9,\;0.95$ and performed extensive simulations.
The difficulty in
this case
lies in the dependence of the ground state energy (and resulting free energy) on the initial distribution of non-magnetic sites. So, for a correctness of results, one should repeat a simulation for a large number of initial configurations and then average the obtained values of the free energy. That is not very practical because it is too time consuming. To accurately estimate the average free energy one needs to increase a number of generated configurations for smaller $p$ because of the distribution width broadening (see Figure \ref{inacc}). It makes this way to be impractical for $p<0.9$.

%
%
%
%
It is possible to overcome this complication. If we take the initial configuration with the ground state energy
value
to be
close enough
to the mean ground state energy for a given value of $p$, the free energy at the critical point on this lattice will be close to the mean value of free energies.
As a result, we generated many lattices with distributed
non-magnetic sites, calculated the ground state energy on these lattices and ran the Wang-Landau algorithm
on the lattice with the energy closest to the mean ground state energy.
Such a way gives
an additional
inaccuracy $\Delta E_{\mathrm{GS}}$ for a free energy value as a difference between the ground state energy of the chosen configuration and the mean one.
This difference becomes irrelevant for large lattices, but noticeable for lattices used in practice (see Figure \ref{inacc}).
We compared these two ways for the case $p=0.9$ and found that they give very similar central values of the central charge.
However, the uncertainty obtained in the second way in our particular simulation was found to be larger by 30\%.

\begin{figure}[t]
   \centering
   \includegraphics[width=0.43\linewidth]{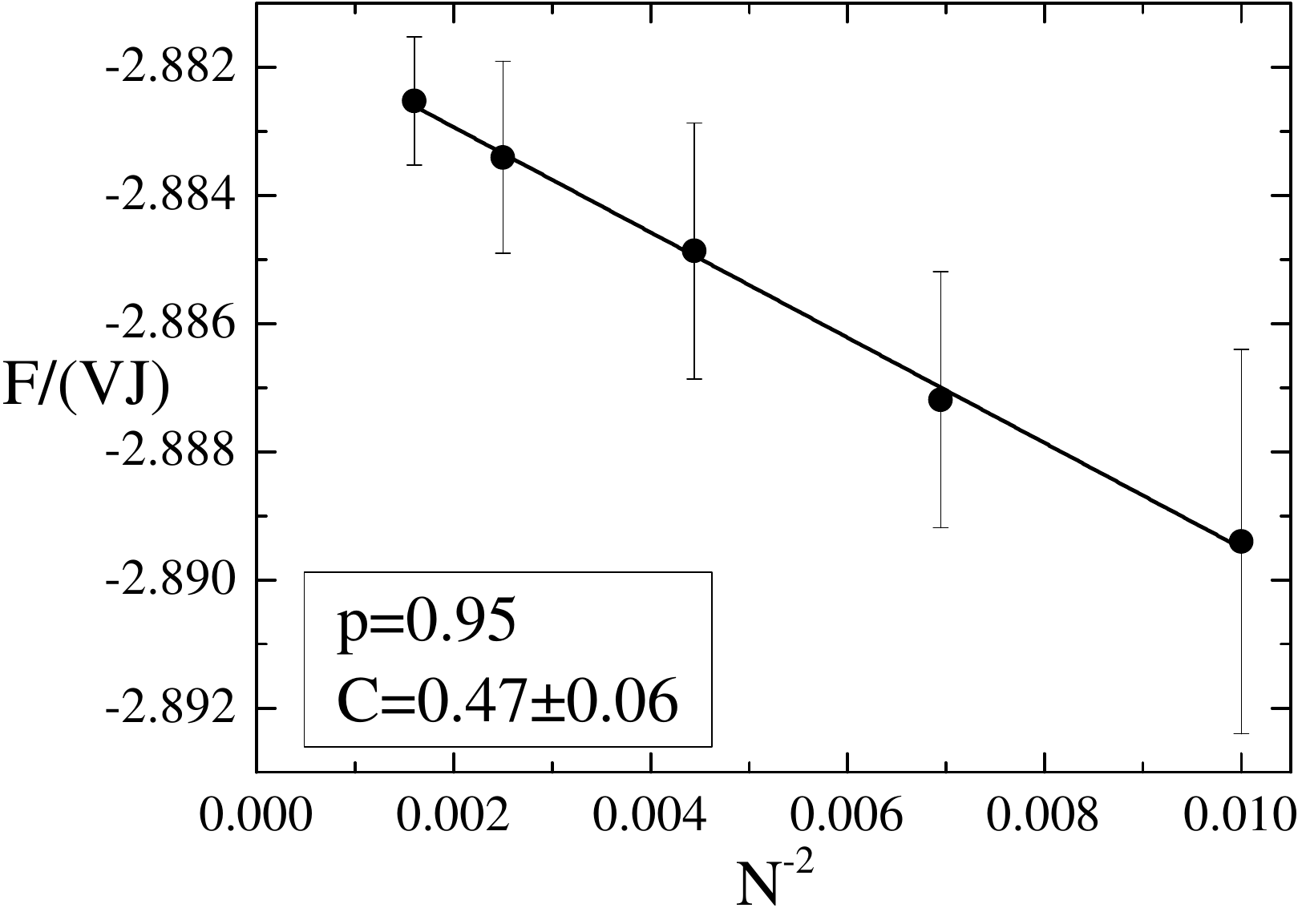}\qquad
   \includegraphics[width=0.43\linewidth]{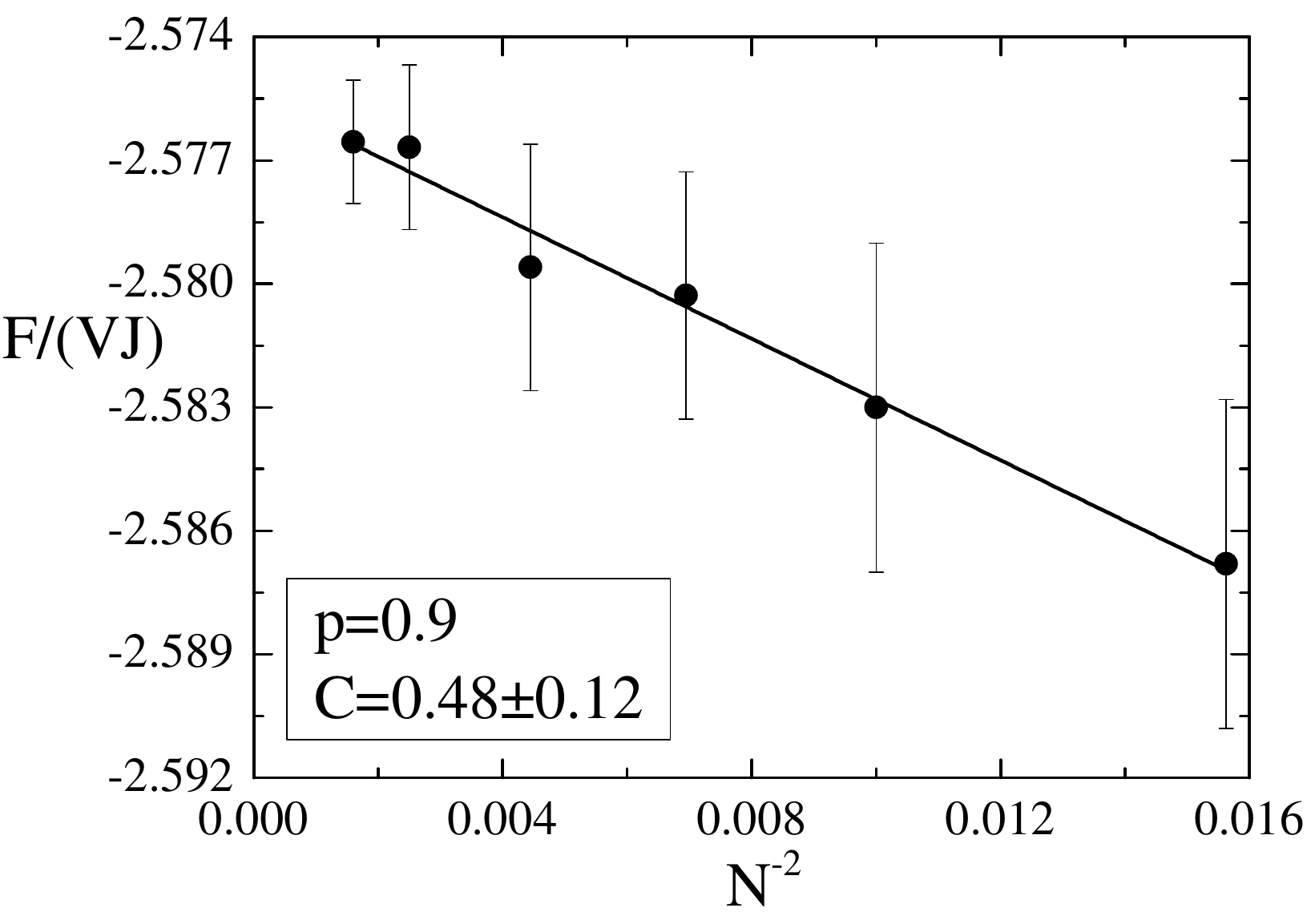}
   \includegraphics[width=0.43\linewidth]{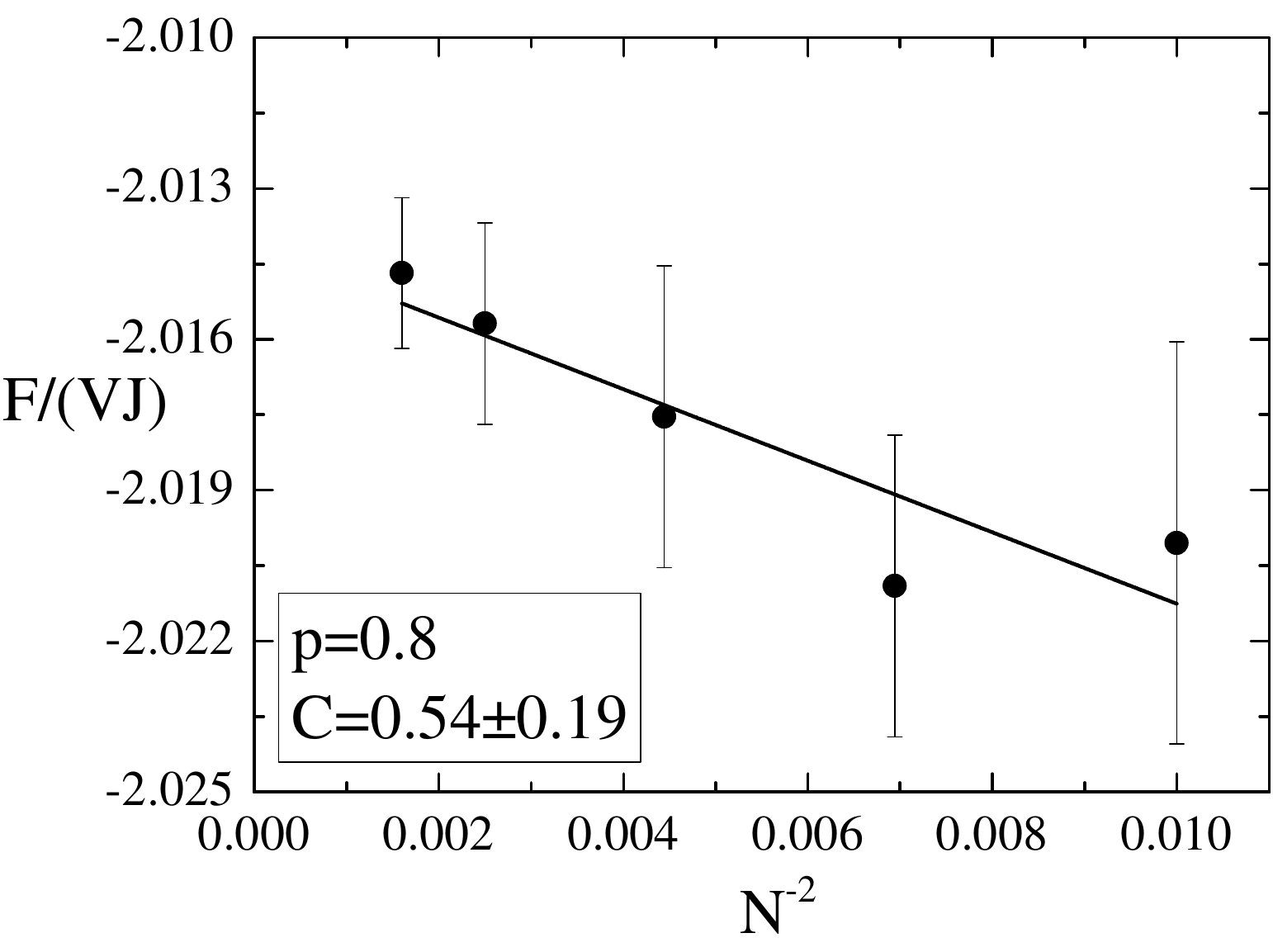}\qquad\quad
   \includegraphics[width=0.39\linewidth]{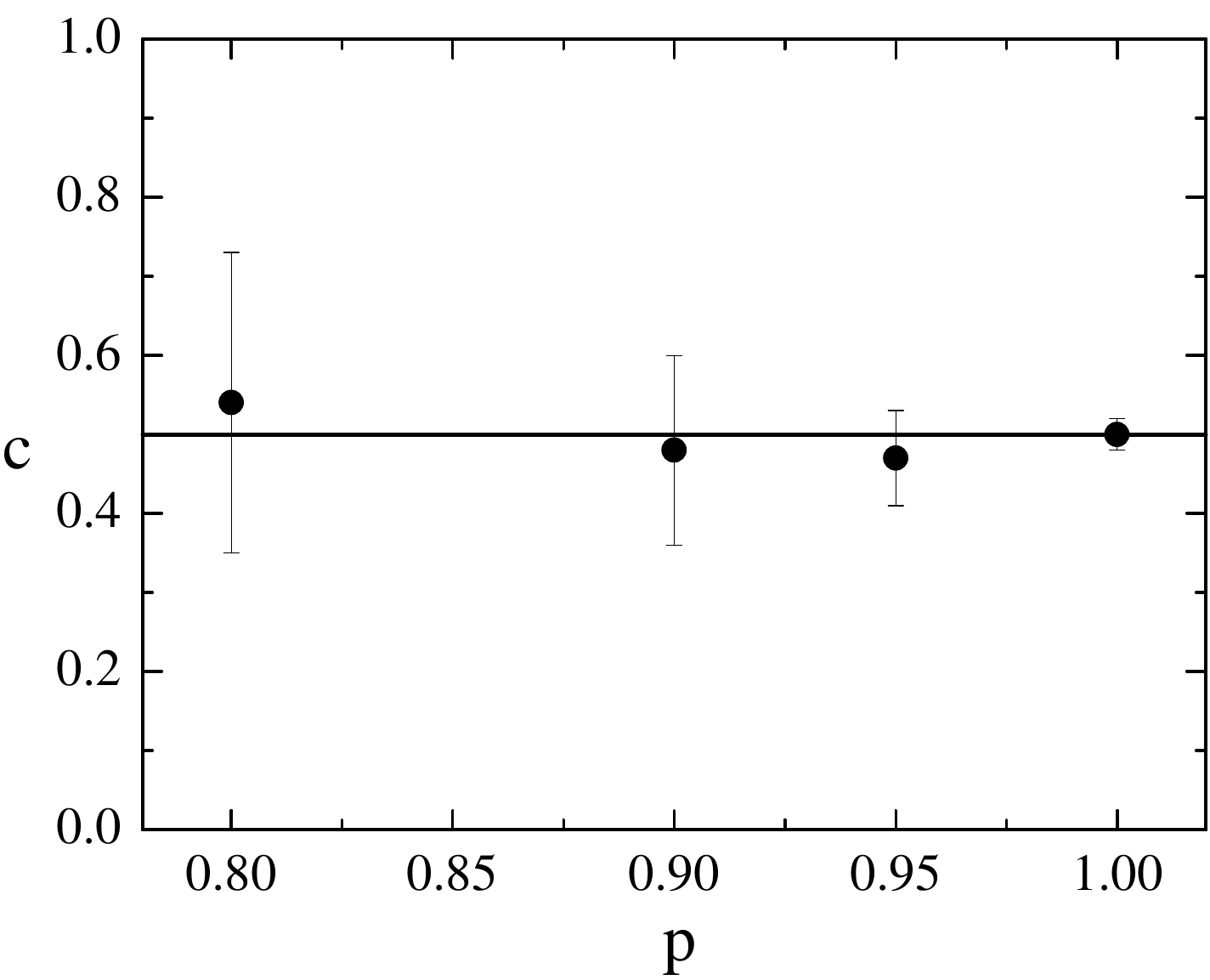}
   \caption{\label{cc09}Estimation of the central charge as a plot slope for the site-diluted Ising model with $p=0.95$ ({\it top left}), $p=0.9$ ({\it top right}) and $p=0.8$ ({\it bottom left}); summary of the result for the central charge values ({\it bottom right}).}
\end{figure}

The results for the central charge for $p\leq 1$ obtained in the second way are shown in Figure~\ref{cc09} and in Table \ref{results}.
In contrast to Ref.~\cite{najafi2016monte}, we see that $c$ remains close to $\frac{1}{2}$ as $p$ decreases, although the inaccuracy becomes larger. 
\begin{table}[h]
  \center
  \begin{tabular}{c|c|c}
     \hline
     \hline
       $p$ & $T_c$ & $c$\\
       \hline
       $1$ & $3.64095\ldots$ & $0.50\pm0.02$ \\
       $0.95$ & $3.368\pm0.002$ & $0.47\pm0.06$ \\
       $0.9$ & $3.084\pm0.003$ & $0.48\pm0.12$ \\
       $0.8$ & $2.499\pm0.005$ & $0.54\pm0.19$ \\
     \hline
     \hline
  \end{tabular}
  \caption{\label{results} Summary of the results for the critical temperature and the central charge.}
\end{table}

This result is confirmed indirectly by
results of the Wolff cluster algorithm~\cite{wolff1989collective,landau2014guide}.
The right plot of Figure~\ref{phase} shows that the values of the critical indices
agree with
those for the pure Ising model.
So, one can expect that for $0.65<p<1$ the site-diluted Ising model has the Ising-like critical behavior with $c=1/2$.
%

\section{Conclusion and discussion}
In this paper, we presented a new method to determine the central charge of a conformal field theory.
We employed our method to the pure two-dimensional Ising lattice model and to the site-diluted Ising model, but it can be applied to other models as well.
We showed that
the site-diluted Ising model has the central charge $c=\frac{1}{2}$ and demonstrates Ising-like behavior for $p<1$. 


The contradiction with the results of Ref.~\cite{najafi2016monte} may stem from the
different approaches.
In our work, we used the global property of the model -- a dependence of the free
energy on the torus modular parameter. The approach of Ref.~\cite{najafi2016monte} is
based on the geometry of domain walls, which may be altered by the appearance of the non-magnetic
sites. We suppose that the different boundary condition changing operator should be used at the tip of the growing domain wall to obtain a correct CFT description. 
The relation between the Schramm-Loewner evolution parameter $\kappa$ and the
central charge $c$, used in Ref.~\cite{najafi2016monte},
is applicable, however, only if the boundary condition changing operator is unaltered \cite{bauer2002sle,bauer2003conformal,bauer2003sle}. So,
for the studied model, the numerical values of $\kappa$
can not be used to determine the central charge.

\section*{Acknowledgments}
\label{sec:acknowledgements}
We are thankful to the organizers of the conference Physica.SPb/2016. 
%
%
Anton Nazarov acknowledges the St. Petersburg State University for a support under the Research Grant No. 11.38.223.2015.
Alexander Sorokin is supported by the RFBR grant No. 16-32-60143.
The calculations were partially carried out using the facilities of the SPbU Resource Center ``Computational Center of SPbU''.

\section*{References}
\bibliographystyle{iopart-num}
\bibliography{listing,bibliography}{} 
\end{document}